\documentclass{aa}  
\usepackage{graphicx}
\usepackage{chemarr}
\usepackage{txfonts}
\usepackage{amsmath}
\usepackage{natbib}

\begin{document}

     \title{A sensitivity study of the neutral-neutral reactions $\rm C + C_3$ and  $\rm C + C_5$ in cold dense interstellar clouds}

     
       \author{V. Wakelam\inst{1,2} \and J.-C. Loison\inst{3} \and E. Herbst\inst{4,5} \and D. Talbi\inst{6} \and D. Quan\inst{7} \and F. Caralp\inst{3} }
     \offprints{V. Wakelam, \email{wakelam@obs.u-bordeaux1.fr}}
     \institute{
      Universit\'e de Bordeaux, Laboratoire d'Astrophysique de Bordeaux, BP89 33271 Floirac Cedex, France \and
      CNRS/INSU, UMR 5804, BP89 33271 Floirac Cedex, France \and
      Institut des Sciences Mol\'eculaires CNRS UMR 5255, Universit\'e Bordeaux 1, 33405 Talence, France \and
       Department of Physics, The Ohio State University, Columbus, OH 43210, USA \and
      Departments of Astronomy and Chemistry, The Ohio State University, Columbus, OH 43210, USA \and Universit\'{e} Montpellier II, Groupe de Recherches en Astronomie et Astrophysique du Languedoc, CNRS, UMR 5024, place Eug\`{e}ne Bataillon, 34095 Montpellier, France
      \and The Chemical Physics Program, The Ohio State University,Columbus, OH 43210, USA
        }


     \date{Received xxx / Accepted xxx }
     
     \abstract
     {}
     { Chemical networks used for models of interstellar clouds contain many reactions, some of them with poorly determined rate coefficients and/or products.   In this work, we report a method for improving the predictions of molecular abundances using sensitivity methods and $ab$ $ initio$ calculations.}
     {Based on the chemical network osu.2003, we used two different sensitivity methods to determine the most important reactions as a function of time for models of dense cold clouds. Of these reactions, we concentrated on those between C and C$_{3}$ and between C and C$_{5}$, both for their effect on specific important species such as CO and for their general effect on large numbers of species.  We then used $ab$ $initio$ and kinetic methods to determine an improved rate coefficient for the former reaction and a new set of products, plus a slightly changed rate coefficient for the latter.}
     {Putting our new results in a pseudo-time-dependent model of cold dense clouds, we found that the abundances of many species are altered at early times, based on large changes in the abundances of CO and atomic C. We compared the effect of these new rate coefficients/products on the comparison with observed abundances and found that they shift the best agreement from $3\times 10^4$~yr to $(1-3)\times 10^5$~yr.}
     {}


     \titlerunning{Sensitivity Study in Cold Interstellar Clouds}
     \authorrunning{Wakelam et al.}

     \maketitle

\section{Introduction}

The chemistry in astrochemical models is driven by large networks including thousands of reactions with poorly determined rate coefficients.  For several years now, astrochemists have been working on the quantification of model uncertainties caused by uncertainties in  rate coefficients.  Using similar methods but different models and networks, \citet{2004AstL...30..566V,2008ApJ...672..629V} and \citet{2005A&A...444..883W,2006A&A...451..551W} have studied the error propagation of such uncertainties in various environments such as dense clouds, diffuse gas, hot cores, and protoplanetary disks.  This approach to quantify the model error is useful quantitatively comparing calculated and observed molecular abundances.  In addition to this important aspect, these authors have developed sensitivity methods for identifying those reactions in complete networks that might be worth studying theoretically or experimentally because model results are very sensitive to their rate coefficients and uncertainties.  Identifying these reactions is critical because detailed experimental or theoretical studies of a single reaction, especially at low interstellar temperatures, can take years.  Guiding chemists and physicist as to which reactions are the most important for them to study is an efficient method for reducing error in chemical models.

"Important" reactions can be defined in several ways.  First, one can define important reactions for specific species; such reactions strongly affect the abundances of individual species through changes in their rate coefficients for formation or destruction processes.  One must be precise with this definition, because even though some reactions dominate the formation or destruction of certain species, changes in their rate coefficients within certain ranges can have little effect. 
\citet{quan2008}, for instance, showed that although molecular oxygen is formed by the reaction $\rm O + OH \rightarrow O_2 + H$, the rate coefficient of this reaction can be changed by several orders of magnitude without affecting the O$_2$ abundance significantly. The reason is that the reaction is also the main route of OH destruction. When the rate coefficient is decreased, the OH abundance increases so that the rate of O$_2$ formation stays constant.  As a counterexample, the reaction $\rm C + C_{3} \rightarrow C_{4} + h\nu$ is very important to the abundance of CO, as will be discussed below. Secondly, "important" reactions can be defined through the number of molecular abundances they influence strongly.  As an example, the cosmic-ray ionization of molecular hydrogen is important for almost all of the species in the chemical models of dense interstellar clouds \citep{2006A&A...459..813W,2008ApJ...672..629V}. 

In this paper, we present a complete approach to improving chemical models through  identifying sensitive reactions followed by an increase in our knowledge of their rate coefficients via theoretical means, and using the newly revised rate coefficients in model predictions.  We direct our focus on two reactions: the radiative association reaction between $\rm C$ and $\rm C_3$ and the neutral-neutral reaction $\rm C + C_5$. We start, in Section 2, with a discussion of the methods used for our sensitivity analyses and the determination of the important reactions in cold dense clouds. The theoretical study of the $\rm C + C_3$ and $\rm C + C_5$ rate coefficients is presented in Section 3 and the impact of new rate coefficients on the model predictions are studied in Section 4. Following our conclusion, an Appendix on other reactions of the $\rm C + C_{n}$ class is presented.

\section{Selection of important reactions}
\label{rx}
In order to identify important reactions of both classes, we use two different sensitivity methods and compare the results of the two. Each sensitivity method is based on linear correlations but in the first one, only one reaction is modified at a time whereas in the second method all reactions are modified at the same time. The first one, which we will call ``individual perturbations'',  is simply based on the study of the linear effect of individual rate coefficient perturbations on molecular abundances.  We have already used this method previously  \citep{2005A&A...444..883W,2006A&A...451..551W}. In the second method, we compute Pearson correlation coefficients.

\subsection{Chemical model}

To represent cold dark clouds, we use a pseudo- time-dependent gas-phase model, which computes abundances as a function of time for a temperature  of 10~K, a total hydrogen density of $2\times 10^4$~cm$^{-3}$ and a  visual extinction of 10. The model is known as nahoon ($\rm http://www.obs.u-bordeaux1.fr/amor/VWakelam/$).  For comparison with previous studies, the elemental abundances are the O-rich low-metal elemental abundances \citep[see table 1 in][]{2008ApJ...680..371W}. For this work, we used an older version of the osu database (osu.2003)  since one goal of this work is to show how we corrected a mistake in this database.  We focus on the specific time of 10$^{5}$ yr, since this so-called ``early time'' normally results in the best agreement between calculated abundances and observations \citep{2004MNRAS.350..323S}.

\subsection{Sensitivity Methods}

The idea of the individual perturbation approach  is to look at the individual effect of each reaction. With a network of $N$ reactions, we run the model  $N$ times and modify only one rate coefficient each run.  One then obtains   the chemical evolution of the fractional abundance: $X^i_j(t)$ for each species $j$ and run $i$ . Note that the index $i$ can stand for the run or the reaction with varied rate coefficient.
We then compare the abundance of each species $X^i_j(t)$ with the standard abundance $X^{ref}_j(t)$ obtained without changing the rate coefficients.  In our previous uses of this approach  to select important reactions for 
hot-core \citep{2005A&A...444..883W} and dense-cloud \citep{2006A&A...451..551W} chemistry,  we multiplied each rate coefficient by its uncertainty factor. This approach assumes that the real uncertainty in the rate coefficient is not much larger than the considered uncertainty factor, which may not be true. Indeed, as we will show later, this approach can miss important reactions if the real rate coefficient error is much larger than the uncertainty factor. For this reason, here we multiply each rate coefficient by the same factor of two. This factor is currently the default uncertainty used for most classes of reactions in current reaction networks (http://www.udfa.net/; \verb+http://www.physics.ohio-state.edu/~eric/research.html+).  To quantify the effect of these variations in the reaction rate coefficients  on the abundances, we compute a parameter labeled $R^i_j(t)$ for each species and each reaction, where
\begin{equation}
\label{req}
R^i_j(t)=\frac{(X^i_j(t)-X^{ref}_j(t))}{X^{ref}_j(t)}.
\end{equation}
As an example, a value of  $R^i_j(t)=0.2$ tells us that the multiplication of the reaction $i$ by a factor of two increases the abundance of species $j$ by 20\%.  We also did these calculations by multiplying the rate coefficients by a factor of 10 or dividing them by 2 but the list of important reactions is rather the same as for the case presented here.

The second method is based on linear correlations but is less time-consuming.  Here we run about 2000 different models $l$, randomly varying all rate coefficients within an uncertainty range of a factor of two, with both increases and decreases considered.  We then compute Pearson correlation coefficients for each species $j$ and each reaction $i$ using the following relation: 
\begin{equation}
P^i_j(t)=\frac{\sum^l (X^l_j(t)-\bar{X_j(t)})(k^l_i-\bar{k_i})}{\sqrt{(\sum^l (X^l_j(t)-\bar{X_j(t)})^2 \sum^l (k^l_i-\bar{k_i})^2}}, 
\end{equation}
where $\bar{X_j(t)}$ and $\bar{k_i}$ are the mean values of the abundance for species $j$ and the rate coefficient $i$. Pearson correlation coefficients indicate the strength and direction of a linear relationship between two variables by dividing the covariance of the two variables by the products of their standard deviations \citep[see for instance][]{NumericalRecipes}.

\subsection{Results of sensitivity calculations}
\label{sens}

In this section, we show the results of both sensitivity methods to determine the most important reactions for the abundances of selected individual species and to determine the reactions that influence the abundances of the largest numbers of  species.


\begin{table}
\caption{Most important reactions for selected species at $10^5$~yr.}
\begin{center}
\begin{tabular}{|l|lcc|}
\hline
\hline
Species & Reaction & R & P \\
\hline
CO & $\rm C + C_3 \longrightarrow C_4 + h\nu $ & 0.009 & 0.5 \\
  & $\rm H_2 + \zeta \longrightarrow H_2^+ + e^- $ & 0.004 & 0.3 \\
 & $\rm C + C_5 \longrightarrow C_6 + h\nu $ & 0.005 & 0.4 \\
  \hline
O$_2$ & $\rm H_2 + \zeta \longrightarrow H_2^+ + e^- $ & 1.2 & 0.7 \\
& $\rm C + C_3 \longrightarrow C_4 + h\nu $ & 0.3 & 0.2 \\
& $\rm H_3^+ + CO \longrightarrow HCO^+ + H_2 $ & -0.3 & -0.2 \\
 & $\rm C + C_5 \longrightarrow C_6 + h\nu $ & 0.2 & 0.2 \\
& $\rm C + O_2 \longrightarrow CO + O $ & -0.2 & -0.2 \\
  \hline
C$_4$H & $\rm O + C_4H \longrightarrow C_3H + CO $ & -0.7 & -0.2 \\
 & $\rm He + \zeta \longrightarrow He^+ + e^- $ & 0.3 & 0.2 \\
 & $\rm C + C_5 \longrightarrow C_6 $ & -0.2 & -0.07 \\
 & $\rm C + C_3 \longrightarrow C_4 $ & -0.2 & -0.08 \\
 & $\rm C_4H_2^+ + e^- \longrightarrow C_4H + H $ & 0.2 & 0.09 \\
   \hline
HC$_3$N  & $\rm He + \zeta \longrightarrow He^+ + e-^- $ & -0.5 & -0.5 \\
 & $\rm H_2 + \zeta \longrightarrow H_2^+ + e^- $ & 0.5 & 0.5 \\  
 & $\rm H_3O^+ + e^- \longrightarrow OH + H + H $ & -0.2 & -0.1 \\
 & $\rm H_3O^+ + e^- \longrightarrow H_2O + H $ & 0.1 & 0.2 \\
 & $\rm H_3^+ + C \longrightarrow CH^+ + H_2 $ & 0.1 & 0.2 \\  
  & $\rm C^+ + H_2 \longrightarrow CH_2^+  $ & 0.1 & 0.2 \\
  & $\rm C^+ + HC_3N \longrightarrow C_3H^+ + CN $ & -0.1 & -0.1 \\
  & $\rm CH_3^+ + H_2 \longrightarrow CH_5^+ $ & 0.1 & 0.1 \\ 
  & $\rm C + C_3 \longrightarrow C_4 $ & -0.1 & -0.09 \\
\hline
\end{tabular}
\end{center}
\label{tabreac1}
\end{table}%


\begin{table}
\caption{Reactions influencing the most species at 10$^{5}$ yr using $R$}
\begin{center}
\begin{tabular}{|lc|}
\hline
\hline
Reaction & \% of species \\
   \hline
$\rm H_2 + \zeta \longrightarrow H_2^+ + e^- $  & 53 \\
$\rm H_3^+ + C \longrightarrow CH^+ + H_2 $ & 31\\
$\rm He + \zeta \longrightarrow He^+ + e^- $ & 31 \\
$\rm C^+ + H_2 \longrightarrow CH_2^+ $ & 29 \\
$\rm H_3O^+ + e^- \longrightarrow H_2O + H $ & 26\\
$\rm HCO^+ + C \longrightarrow CH^+ + CO $ & 24\\
$\rm CH_3^+ + H_2 \longrightarrow CH_5^+ $ & 22\\
$\rm HCO^+ + C_3 \longrightarrow C_3H^+ + CO $ & 16\\
$\rm C + C_3 \longrightarrow C_4 $ & 15 \\
$\rm H_3^+ + CO \longrightarrow HCO^+ + H_2 $ & 14 \\
$\rm C_3^+ + e^- \longrightarrow C_2 + C $ & 13 \\
$\rm H_3O^+ + e^- \longrightarrow OH + H + H $ & 13\\
$\rm CH_5^+ + CO \longrightarrow HCO^+ + CH_4$ & 13\\
   \hline
\end{tabular}
\end{center}
\label{list_specR}
\end{table}%

\begin{table}
\caption{Reactions influencing the most  species at 10$^{5}$ yr using $P$}
\begin{center}
\begin{tabular}{|lc|}
\hline
\hline
Reaction & \% of species \\
   \hline
$\rm H_2 + \zeta \longrightarrow H_2^+ + e^- $  & 86 \\
$\rm He + \zeta \longrightarrow He^+ + e^- $ & 82 \\
$\rm H_3^+ + O \longrightarrow OH^+ + H_2 $ & 54 \\
$\rm H_3O^+ + e^- \longrightarrow OH + H + H $ & 48\\
$\rm H_3^+ + CO \longrightarrow HCO^+ + H_2 $ & 43 \\
$\rm C + C_5 \longrightarrow C_6 $ & 41 \\
$\rm C + C_3 \longrightarrow C_4 $ & 41 \\
$\rm C^+ + H_2 \longrightarrow CH_2^+ $ & 41 \\
$\rm H_3O^+ + e^- \longrightarrow H_2O + H $ & 37\\
$\rm HCO^+ + e^- \longrightarrow CO + H $ & 34\\
$\rm CH_3^+ + H_2 \longrightarrow CH_5^+ $ & 33\\
$\rm H_3^+ + C \longrightarrow CH^+ + H_2 $ & 32\\
$\rm HCO^+ + C \longrightarrow CH^+ + CO $ & 24\\
   \hline
\end{tabular}
\end{center}
\label{list_specP}
\end{table}%

In Table~\ref{tabreac1}, we show the most important reactions for the species CO, O$_2$, C$_4$H and HC$_3$N at an integrated time of $10^5$~yr using the individual perturbation ($R$) and the correlation coefficient ($P$) methods. In tables~\ref{list_specR} and \ref{list_specP}, we list the reactions that influence the most species using $R$ and $P$, respectively, as a criterion. For each reaction, we count the number of species for which  
$R$ or $P$ is either larger than 0.1 or smaller than $-0.1$. 

A glance at Table~\ref{tabreac1} shows that the important reactions governing the abundances of individual species are generally similar for both methods.  There are two main surprises, however.   First, for the case of CO, the values of $R$ are much smaller than those for $P$.  Presumably, the explanation for this dichotomy is that most of the carbon goes into CO at that time. As a result, a change in the rate coefficients does not change the CO abundance, which is then controlled by the elemental abundance of C. This effect also explains the small uncertainty in the CO abundance found by \citet{2006A&A...451..551W}.   Secondly, the most unexpected reactions, because they seem so unrelated to the listed species, are those between C and $\rm C_3$ and between C and $\rm C_5$.  Note that the latter was assumed, erroneously, in osu.2003 to be a radiative association (see below).  The importance of the C + $\rm C_3$ association seems to be associated with the fact that C$_{3}$ acts as a catalyst to convert C and O into CO via the association reaction between C and C$_3$ followed by

\begin{equation}
{\rm O + C_{4} \longrightarrow CO + C_{3},}
\end{equation}
which regenerates the C$_{3}$.  A similar cycle happens with C$_{5}$ if it undergoes radiative association with C.

Considering the number of species affected, the most important reactions determined by both methods are quite similar although the order of importance can differ.   As already noticed by several authors \citep{2006A&A...451..551W,2004AstL...30..566V,2006A&A...459..813W,2008ApJ...672..629V}, the H$_2$ and He ionization rates by cosmic-rays  appear to be important for many species since they are at the origin of formation and destruction pathways. Most of the other important reactions are ion-neutral normal and associative reactions and  dissociative recombination reactions, such as those for H$_3$O$^+$ with electrons.  Among the important reactions, $\rm C^+ + H_2 \longrightarrow CH_2^+ $, $\rm CH_3^+ + H_2 \longrightarrow CH_5^+ $, $\rm H_3^+ + C \longrightarrow CH^+ + H_2 $, $\rm H_3^+ + CO \longrightarrow HCO^+ + H_2 $, $\rm H_3^+ + O \longrightarrow OH^+ + H_2 $ have also been identified as important reactions for disk chemistry by \citet{2008ApJ...672..629V}. Most of these reactions are at the start of a chain of reactions that builds up hydrocarbons. It should be emphasized that our sensitivity methods need not tell us the
details of the pathway of reaction cycles.  If, for example, a destruction
reaction leads to a loop of subsequent reactions that reforms the molecule
being destroyed with some non-zero efficiency, the sensitivity method might
just decide that the initial destruction reaction is less important, at
least for the species being destroyed.   The problem of enhanced molecular
abundances due to artificial loops has been minimized whenever possible in
the construction of the network by including a number of different product
channels.  For example, if the major destruction route of a neutral species
leads to a protonated ion, and the major destruction route of the ion is
dissociative recombination to form neutral molecules, the channel that loops
back to the original neutral species plus a hydrogen atom will not be
assumed to occur on more than 50\% of the recombination reactions (unless, as
is highly unusual, experimental results tell us the contrary).  Let us look
at the important reactions causing the destruction of HC$_3$N in Table~\ref{tabreac1}.  One
of them is the rather direct destruction via C$^+$ ions.  The products of this
reaction do not lead back to HC$_3$N in any direct fashion.  If, on the other
hand, there were a charge exchange reaction to form C$_3$HN$^+$, and this ion were
hydrogenated by reaction with H2 to form, C$_3$H$_2$N$^+$ + H, the hydrogenated ion
could react with electrons to form HC$_3$N + H.  In the network, however, the
probability of this neutral product channel is set to approximately 50\% so
that the loop is not particularly efficient.

 The only neutral-neutral reactions that seem important for many species in Tables~\ref{list_specR} and \ref{list_specP} are the (presumed) radiative associative reactions $\rm C + C_3 $ and $\rm C + C_5$. The rate coefficients for these reactions used in osu.2003 are $10^{-10}$ and $10^{-11}$~cm$^3$~s$^{-1}$ based on \citet{2004MNRAS.350..323S}, without temperature dependence.  These reactions have been looked at again by \citet{2006smith} and it was realized that the reaction between C and $\rm C_5$ has the normal exothermic products $\rm C_3 + C_3$.  Given the importance of these reactions, we decided to study them theoretically in some detail.

\section{New calculations for C + C$_3$ and C + C$_5$ reactions}

\subsection{C + C$_{3}$}

At low collision energies, the reaction of C($^3$P) atoms with linear C$_3$ in its ground electronic state  $(\rm X~^{1}\Sigma^{+}_{\rm g})$ leads only to adduct formation, C$_4$, because the exit channel C$_2$ + C$_2$ is endothermic. The lowest energy pathway to produce C$_{4}$ is the linear one, and we focus on this pathway here. Under  interstellar low-density conditions, the only way to stabilize the C$_4$ adduct is by emission of a photon. The formation of the linear C$_4$ adduct is strongly exothermic      and spin-allowed to its triplet manifold of states, designated $^{3}\rm C_{4}$: 
\begin{equation}
\rm C(^3P) + ^1C_3 \longrightarrow ^{3}C_4 + h\nu .    \qquad    
\label{c4}         
\end{equation}
There are a variety of values for this exothermicity.   A high-temperature experiment yields a value of  (5.21 $\pm 0.21$) eV from the heats of formation of C$_{4}$ and C$_{3}$ \citep{Gingerich1994}, a PST calculation based on a photodissociation experiment yields a lower value of (4.84 $\pm 0.15)$ eV \citep{Choi2000}, while high-level $ab$ $initio$ theoretical values (including our own work) computed at the RCCSD and RCCSD(T) levels range from 4.94-4.98 eV without zero-point correction  \citep{Martin1995}.   The high-temperature experimental value has been used in the RRKM calculations discussed below. The rhombic isomer of C$_{4}$ in its ground singlet state has been calculated to to be nearly isoenergetic with the linear isomer \citep{2006Masso},  but the pathway from C and linear 
C$_{3}$ to the first excited triplet state shows a high barrier at the CASSCF level of $ab$ $initio$ theory (see below).

In the $\rm C_{\infty,v}$ symmetry, the three electronic states of C$_4$ correlating to the three spin-orbit states of C($^3$P$_{0,1,2}$) lead to two different potential energy surfaces corresponding to the fundamental state $\rm X~^3 \Sigma^-_g$ and the first excited $^3\Pi _g$  state. (Note that the ground electronic state of C is of g symmetry.)  The energetic and structural parameters of these states of C$_{4}$ and the second excited $^3\Pi_u$ state are listed in Table~\ref{table2}.  These parameters and others such as the pathway to the rhombic isomer of C$_{4}$  have been calculated at the Restricted Hartree-Fock (RHF) and Complete Active Space Self Consistent Field (CASSCF) level with the double-zeta cc-pVDZ basis set of \citet{1989Dunning}, using MOLPRO 2000, a package of $ab$ $initio$ programs designed by H.-J. Werner, and P. J. Knowles (see http://www.molpro.net for more details).  This level of theory was also used to 
 determine if the three electronic curves are attractive along the linear reaction pathway. 
 In these calculations, which underestimate the exothermicity of reaction~({\ref{c4}),  the C$_{\rm \infty v}$ (linear) approach of C toward the C$_3$ axis, shown in Figure~\ref{fig1}, yields no barrier for the ground C$_4$($^3 \Sigma ^-_g$) electronic state but shows a high barrier of 1.5 eV for the first excited C$_4$($^3\Pi_{\rm g}$) state.   Using $\Omega$ and $\rm M_J$ conservation, we obtain that only the C($\rm ^3P_0$) and C($\rm ^3P_1$, $\rm M_J=-1,+1$) fine-structure substates are reactive; these three substates form  vibrationally excited linear C$_4$ in its fundamental $^3 \Sigma^-_g$ state, while the other six substates lead to the formation of the $\rm ^3\Pi_g$ excited state (through a high barrier). Any others approach on the triplet surface, as the perpendicular one leading to the rhombic C$_4$, shows high barriers at the CASSCF/cc-pVDZ level.  As the C$_4$ is formed with a great excess of energy,  the potential minima of several electronic states, not just the ground state, lie below the dissociation limit (the potential energy of ground-state reactants.). Stabilisation of C$_4$ adduct can then occur via purely vibrational transitions from the ground-state adduct, but also via electronic transitions from excited state adducts if they can be formed without barriers \citep{1988ApJ...329..410H}. Here, we must therefore consider three mechanisms for adduct stabilisation:
\begin{itemize}
\item  vibrational radiative transitions from quasi-continuum vibrational levels in the ground state 
$\rm (X~^3\Sigma^-_g)$ adduct to lower bound vibrational levels,
\item   ``inverse internal conversion'' from the $X~^3 \Sigma^-_g$ adduct to an excited electronic state followed by electronic photon emission to bound vibrational levels of the ground state (case 1),
\item   direct radiative emission from  quasi-continuum vibrational levels in the ground  $\rm X~^3 \Sigma^-_g$ state to lower vibrationally bound levels in excited states (case 2).
\end{itemize}

Other possibilities arise from the isomerisation of the triplet linear C$_4$ state toward the higher-energy singlet rombric C$_4$ state. However this isomerisation has to overcome a high barrier as well as intersystem crossing. We can reasonably neglect those possibilities.

\begin{table}
\caption{Vdz/CASSCF optimized structural parameters
 }
\begin{tabular}{|c|c|c|c|c|c|c|}
\hline
\hline
State & T$_0$ (eV) th/exp & R(C$_{1}$-C$_{2}$) & R(C$_{2}$-C$_{3}$) & R(C$_{3}$-C${4}$) & $\theta 1$ &  $\theta 2$ \\
\hline
$^3\Pi _u$ & 0.89/0.93 & 1.27 & 1.33 & 1.27 & 180 & 180 \\
$^3\Pi _g$ & 0.63/0.82 & 1.28 & 1.34 & 1.28 & 180 & 180 \\
$^3\Sigma _g^{-}$ & 0 & 1.32 & 1.30 & 1.32 & 180 & 180 \\
\hline
\end{tabular}
\label{table2}
\\
Distances $R$ in $\AA$, angles in degrees and total energies are with respect to the $\rm X^3 \Sigma^-_g$ state. The results presented in the table are from this work except the experimental data, which are from \citet{2000Linnartz} for $\rm ^3\Pi_g$ and \citet{1997Xu} for $\rm ^3\Pi_u$.
\end{table}%

\begin{figure}
\includegraphics[width=1\linewidth]{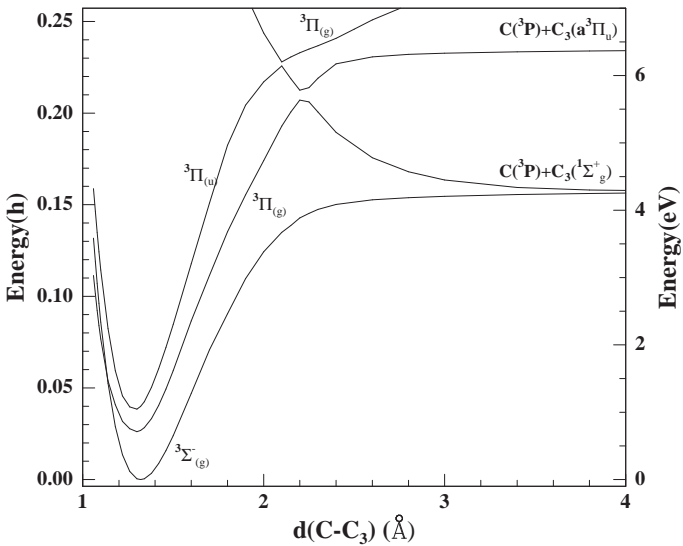}
\caption{Co-linear VDZ-CASSCF one-dimensional cuts of the lowest triplet electronic states of linear C$_4$ molecules along the C$_3$-C stretch. The C$_3$ bonds are optimized. The labeling of the C$_{4}$ electronic states is based on D$_{\rm \infty h}$ symmetry.  \label{fig1}}
\end{figure}

Of the three itemized mechanisms, the case 2 situation can be ignored.  Both experiments and detailed calculations on our part show that the potential surfaces of all relevant low-lying excited states of linear C$_4$ lie almost always above the ground state at all relevant geometries so that emission transition from the ground state are impossible or very inefficient. 
Therefore, the only possibility for stabilization via electronic emission is case 1, with the dominant channel being potential curve crossing from  $\rm X~ ^3 \Sigma ^-_g$ to $^3\Pi _u$  at long C$_2$-C$_2$ distances \citep[see][ for example]{2006Masso} followed by an allowed u$\rightarrow$g electronic transition.  The u-g potential curve crossing is not allowed when C$_{4}$ lies in the symmetric  D$_{\rm \infty h}$ linear configuration, but is allowed by vibrational excitation,  which reduces the symmetry.
Since the $^3\Pi _u$ state has an electronically allowed transition not only with the fundamental state 
$\rm X~^3 \Sigma^-_g$ but also with the first electronically excited $^3\Pi _g$ state, we have to include these two transitions.

If we designate unstable adducts (``complexes'') with an asterisk,  the overall mechanism for association can now be written as:
\begin{equation}
\rm C + C_3 \xrightarrow{k_{ass}} (C_4(^3\Sigma^-_g))^*  \xrightarrow{k_{diss}} C+ C_3
\end{equation}
\begin{equation}
\rm (C_4(^3\Sigma^-_g))^* \xrightarrow{k_{rad,v}} C_4(^3\Sigma^-_g) + h\nu
\end{equation}
\begin{equation}
\rm  (C_4(^3\Sigma^-_g))^*\xrightleftharpoons[k_{-cc}]{k_{cc}} C_4(^3\Pi_u, ^3\Pi_g)^*
\end{equation}
\begin{equation}
\rm C_4(^3\Pi^-_u)^* \xrightarrow{k_{rad,el}} C_4(^3\Sigma^-_g, ^3\Pi_g) + h\nu,
\end{equation}
where ass and diss designate association and dissociation, rad,v designates vibrational emission, cc designates curve crossing, rad,el designates electronic emission,  and $k$ designates a rate coefficient. Assuming that curve crossing is more rapid than the assorted emission processes,  we obtain the following expression for the rate coefficient of the overall radiative association (k$_{\rm RA}$):
\begin{equation}
 k_{\rm RA} =  k_{\rm ass} [k_{\rm rad,v} + k_{\rm rad,el}K_{\rm cc}] / (k_{\rm rad,v} + k_{\rm diss}) \qquad,
\end{equation}
where the equilibrium coefficient $K_{\rm cc}$ is equal to $k_{\rm cc}/ k_{\rm -cc}$ and pertains to the
$^3\Pi_u$ only .

We first determine  the microcanonical rate coefficient for  dissociation, $k_{\rm diss}$, from the ground electronic state of linear 
C$_{4}$ as a function of internal energy $E$ above the C + C$_{3}$ dissociation energy.  We employ the so-called RRKM (Rice-Ramsperger, Kassel, \& Marcus) unimolecular approach \citep{1996Holbrook} using the FALLOFF system  \citep{1993Forst}.  A so-called loose transition state is used with the canonical flexible transition state theory (CFTS) approach \citep{1996Holbrook}.  The needed structural and thermochemical parameters of the studied system derive from quantum calculations.  
Thermal averaging of the microcanonical rate coefficient for dissociation of the C$_{4}$ adduct obtained from the RRKM calculation allows us to calculate the thermal dissociation rate coefficient.  The RRKM thermally averaged lifetime of the C$_{4}$ adduct,  $\rm \tau = \frac{1}{k_{diss}(T)}$,  is presented in Fig.~\ref{fig2} as a function of $T$.  From the equilibrium constant governing formation and dissociation of the adduct, the rate coefficient for formation, which differs from $k_{\rm ass}$ in that fine-structure correlations are not yet taken into account, can be obtained.

\begin{figure}
\includegraphics[width=1\linewidth]{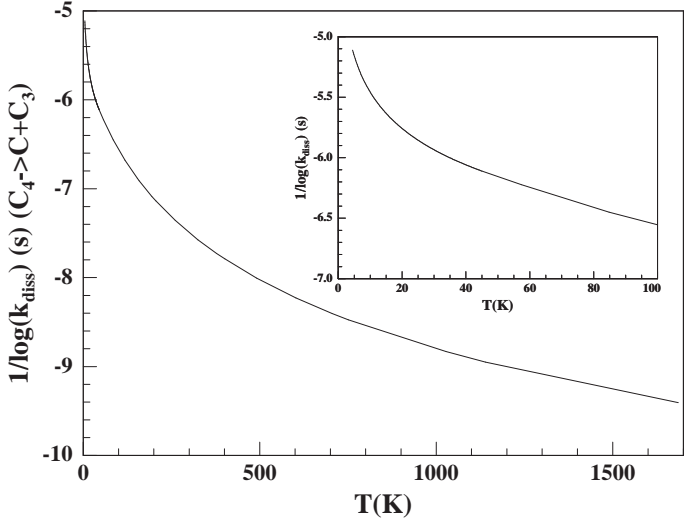}
\caption{RRKM lifetime (in s) of the C$_4$ adduct as a function of the internal energy (in K) above the C + C$_3$ dissociation limit.  \label{fig2}}
\end{figure}


The calculated capture rate constant obtained by this procedure at 298~K is equal to $2.0 \times 10^{-10}$~cm$^3$ s$^{-1}$ with very little temperature dependence in the 10-300 K range,  in good agreement with theoretical treatments for reactions between atomic carbon and assorted alkenes and alkynes \citep{1994ApJ...422..416C} and with the experimental results of \citet{2001A&A...365..241C} for such reactions.  
This thermal rate coefficient, however, does not take into account the fraction of fine-structure states leading to the attractive ground state C$_{4}$ surface.  With this factor and the assumption that the fine-structure states are in thermal equilibrium, we obtain a refined association rate coefficient of  $f(T)$ $\times$ 2 $\times 10^{-10}$~cm$^3$ s$^{-1}$ in the range 10-1000~K, where
\begin{equation}
\label{finestruc}
 f(T) = (1+2\rm e^{-16.4/T})/(1 + 3\times e^{-16.4/T} +  5\times e^{-43.4/T}) .
 \end{equation}

The method of estimating $\rm k_{rad,v}$  equates this rate coefficient with the overall vibrational emission rate  \citep{1982Herbst,1989Smith}:
\begin{equation}
 A(E_{\rm vib}) = \sum_i \sum_{n_i} P_{n_i}^{(i)} (E_{vib})A_{n_i \rightarrow n_i -1}^{(i)}
\end{equation}
where $A_{n_i \rightarrow n_i -1}^{(i)}$,  the Einstein $A$ coefficient for mode $i$ spontaneously emitting from level n$_i$ to n$_i$-1, is  equal to $\rm n_i \times A_{1 \rightarrow 0}^{(i)}$ ), $\rm P_{n_i}^{(i)} (E_{vib})$  is the probability of finding mode $i$ in level n$_i$, and the sums are over all modes $i$ and all energetically accessible quanta n$_i$. The vibrational frequencies and Einstein coefficients, calculated at the B3LYP/vtz level, are listed in Table~\ref{table2}, leading to $k_{\rm rad,v}$  = 349 s$^{-1}$.

\begin{table}
\caption{Vibrational properties of the  $\rm X~^3 \Sigma ^-_g$ state of the C$_4$ molecule. }
\begin{center}
\begin{tabular}{l|c|c|c|c}
\hline
\hline
& $\nu$ (cm$^{-1}$) & IR int & A$_{1,0}$ (s$^{-1}$) & $\rm \sum n_i A_{1,0}$ (s$^{-1}$)  \\
& & (km.mole$^{-1}$) & & \\
\hline
v1 & 172 & 27 & 0.01 & 0.34 \\
v2 & 172 & 27 & 0.01 & 0.34 \\
v3 & 352 & 0 & 0 & 0 \\
v4 & 352 & 0  & 0 & 0 \\
v5 & 908 & 0 & 0 & 0 \\
v6 & 1548 & 312 & 89 & 342 \\
v7 & 2032 & 0 & 0 & 0 \\
\hline
\end{tabular}
\end{center}
\label{table1}
\end{table}%

To estimate the rate coefficient $k_{\rm rad,el}$ for stabilization by emission of electronic radiation, we use the results  of $ab$ $initio$ calculations that the electronic potential curves of the C$_4$($^3\Pi_u$),  C$_4$($^3\Sigma^-_g$) and C$_4$($^3\Pi_g$)  states have very similar shapes.  Therefore, we can approximately ignore nondiagonal Franck-Condon factors.  With the additional assumption that the diagonal  factors are unity, we can write that $<\nu'|\nu''>=\delta_{\nu',\nu''}$, so that the spontaneous electronic emission rate coefficients are given by the expression (esu-cgs units): 
\begin{equation}
k_{\rm rad,el} = (64\pi^4e^2a_0^2/3h)\nu^3|\mu^{el}|^2,
\end{equation}
where $e$ is the electronic charge, $\nu$ is the transition frequency, $a_0$ is the Bohr radius, and  $|\mu^{el}|$ is the electronic transition moment. From the $^3\Pi _u$ state, there are allowed transitions toward the $^3\Sigma^-_g$ and the 
$^3\Pi_g$ state \citep{1983Larson}.  We obtain that  $k_{\rm rad,el} \approx 6\times 10^4$ s${-1}$.  
Finally, we obtain the equilibrium coefficient $K_{\rm cc}$ by assuming that the relative populations of the 
 $^3\Pi_u$, $^3\Pi_g$, and $^3\Sigma^-_g$ states are equal to their relative densities of vibrational states  at an internal C$_{4}$ energy of 5.21 eV, the experimental exothermicity of reaction~(\ref{c4}).   
 This assumption leads to an equilibrium coefficient $K_{\rm cc}$ for the $^3\Pi_u$ state only of 0.15, and an overall  radiative stabilization rate   $k_{\rm rad,v} + k_{\rm rad,el}K_{\rm cc}$ of  9350 s$^{-1}$.  
 
 Once the partial rate coefficients are all calculated, we can obtain the total rate coefficient for radiative association, $k_{\rm RA}$ as a function of temperature.  The results are shown in Figure~\ref{fig3}; the value at 15 K is $ 3 \times 10^{-12}$ cm$^3$ s$^{-1}$.  The value at 15 K if only vibrational stabilization is allowed is a much lower $1 \times 10^{-13}$ cm$^{3}$ s$^{-1}$, which serves as a reasonable lower limit.   The value obtained with the  $<\nu'|\nu''>=\delta_{\nu',\nu''}$ assumption
is certainly an extreme upper limit, because the diagonal Franck-Condon overlaps are certainly lower than unity,  the electronic transition moment, assumed to be independent of energy, may well decrease with increasing energy for certain states \citep{2002Boye}, and the internal conversion from the ground $X~^3\Sigma^-_g$ of C$_{4}$ to excited electronic states may not reach equilibrium.  Estimation of the size these factors have on the association rate coefficient is not facile, but a reasonable  guess is that they contribute to produce a factor of $\approx 0.25$, leading to a reduced value at 15 K of  $8 \times 10^{-13}$ cm$^{3}$ s$^{-1}$.  We adopt as an error limit an uncertainty of $\pm$ 75\%, allthough this error does not quite cover the extreme maximum and minimum values discussed above.  From the RRKM calculations, we obtain a temperature dependence for $k_{\rm RA}$ of  approximately $T^{-1}$, in excellent agreement with the approach of  \citet{1988rcia.conf...17B} for a system with two rotational degrees of freedom. We obtain finally a temperature-dependent association rate coefficient $k_{\rm RA}(T)$ of $(4.0 \pm 3.0) \times 10^{-14} \times (T/300)^{-1}$ cm$^{3}$ s$^{-1}$ in the range 10-300 K.    Compared with the radiative association rate coefficient of $1 \times 10^{-10}$ cm$^{3}$ s$^{-1}$ in \citet{2004MNRAS.350..323S} and osu.2003, our new value is lower by  a factor of $\approx$ 100 at 10 K.  This discrepancy between old and new values is larger than the formerly assumed uncertainty of an order of magnitude.  

\begin{figure}
\includegraphics[width=1\linewidth]{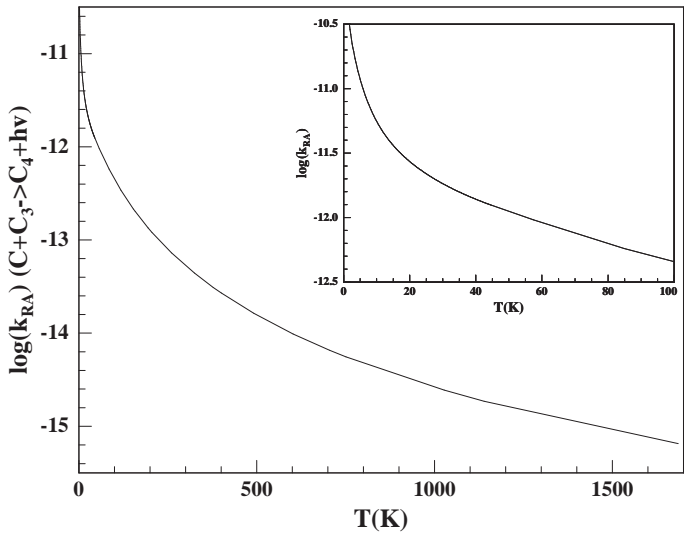}
\caption{Radiative association rate coefficient for C$_{4}$ production as a function of  temperature.  \label{fig3}}
\end{figure}


\subsection{ C + C$_{5}$}

For the reaction between C in its ground $^3\rm P$ state and  linear C$_{5}$ in its ground $^1\Sigma^+_g$ state, the formation of the linear C$_6$ adduct in its $^3\Sigma^-_g$ ground state is strongly exothermic and spin-allowed. This case is then very similar to the C + C$_3$ one, and to determine if the electronic curves are attractive along the entrance channel reaction pathway, we performed similar $ab$ $initio$ calculations.  Our results are also similar, with no barrier for the C$_6$($^3\Sigma^-_g$) electronic state but high barriers for the C$_6$($^3\Pi$) electronic states.  Unlike the case of the C$_{4}$ adduct, however, the C$_6$ adduct can fragment by an exothermic route; in particular
\begin{equation}
\rm	C(^3P)  +  C_5(^1\Sigma +_g) \longrightarrow  C_6(^3\Sigma -_g) \longrightarrow  C_6(^1\Sigma) \longrightarrow  C_3(^1\Sigma +_g) +  C_3(^1\Sigma +_g)  \qquad  ,
\end{equation}
with an overall reaction enthalpy of $\Delta H$  = -1.36 eV.  Although this reaction is formally spin-forbidden to the products in their singlet ground states, the adduct is likely to last long enough to allow intersystem crossing, as in the C + C$_2$H$_2$ reaction \citep{2007Mebel}, and the barrier on the singlet exit channel is certainly much lower than the exothermicity \citep{1982Nelson}. The first excited triplet state of C$_{3}$ lies about  2 eV above the ground singlet state so is probably not accessible. Based on experiments, the assumption that the rate coefficient is independent of the number of hydrogen atoms on the C$_{5}$ reactant, and using the fine structure factor in eq.~(\ref{finestruc}) , we estimate the rate coefficient to form the adduct and subsequent products to be $1.2 \pm(0.8) \times 10^{-10}$ cm$^3$ s$^{-1}$ for 10-300 K. This value is somewhat less than that obtained from the capture methodology of \citet{1994ApJ...422..416C}.   Compared with the value in osu.2003 ($k_{\rm RA} = 1.4 \times 10^{-10}$ cm$^{3}$ s$^{-1}$),  the rate coefficient for the reaction $\rm C + C_5$ is approximately the same but the products are C$_3$ + C$_3$ instead of C$_6$.

The rate coefficients for other $\rm C + C_n$ reactions are discussed in the Appendix. These reactions are not among those  the model results are very sensitive to.

\section{Impact on predicted abundances}

\subsection{General impact}

\begin{figure}
\includegraphics[width=0.8\linewidth]{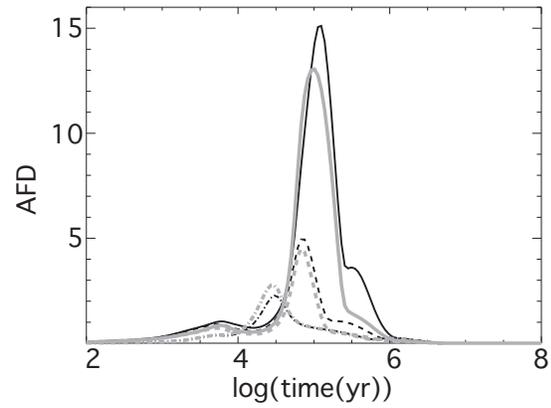}
\caption{Average fractional difference (AFD, see text) as a function of time. The dashed lines are for case (1), the dashed-dotted lines for case (2), and the solid lines for case (3).  
Black lines are for case (a) whereas grey curves are for case (b).
\label{SFR}}
\end{figure}

To quantify the overall impact of change in the rate coefficients/products computed in the previous section, we sum $R^i_j(t)$,  the  fractional difference for species $j$ and reaction $i$ as defined in equation~(\ref{req}),  over the species $j$ and divide by the number of species $n_{\rm j}$ to obtain the average fractional difference $AFD(t)$ for reaction $i$: 
\begin{equation}
 AFD(t) = \frac{1}{n_j} \sum_j |R^i_j(t)|.
\end{equation}
Here, as in equation~(\ref{req}), we define the reference fractional abundance $X^{ref}_j(t)$ for each species as that obtained with the old rate coefficient/products, but define the new values of $X^i_j(t)$ as those obtained with the modified values for reactions $i$ obtained here.
 An unimportant reaction at a specific time would give an $AFD$ close to zero at that time.  It should be noted that although we are using a method similar to section 2.2, the goal here is completely different. In the sensitivity analysis section (2.2), we multiply all reactions, one after another by a factor of 2 in order to compare the abundance sensitivity to individual reactions. Here, we change the rate coefficients for which new calculations have been done, for instance the C+C$_3$ reaction is changed by two orders of magnitude, and we look how much the new rate coefficients have changed the model predictions. If we redo the sensitivity analysis as in section 2.2 with the updated network, we may find different "important" reactions since the network has been changed.
 
For the dense cloud model discussed in Section~\ref{rx}, we consider three different cases for reactions $i$: (1), we change only the rate coefficient of the C+C$_3$ reaction, (2), we change only the products and the rate coefficient of the C+C$_5$ reaction and (3), we perform the changes for both reactions. For each of these cases, we compute the $AFD$ parameter by (a) summing over all species (452),  and by (b) summing only over those species with a fractional abundance greater  than $10^{-10}$ at $10^5$~yr.  For (b), the sum is made over 79, 80 and 77 species for cases (1), (2) and (3) respectively.   
Fig.~\ref{SFR} shows  the results for the six different cases.  It is clear from the results that the most important effect is obtained by the combined effect of changing both C+C$_3$ and C+C$_5$ reactions. For case (3), the sum over all species yields an AFD of 15 at slightly more than 10$^{5}$ yr, while the sum over the most abundant species only yields an AFD of 13 at $10^5$~yr. If only one of the two reactions is changed, C+C$_3$ has the greater effect whereas C+C$_5$ affects the species at early times.

\subsection{Impact on specific species}

\begin{figure}
\includegraphics[width=0.8\linewidth]{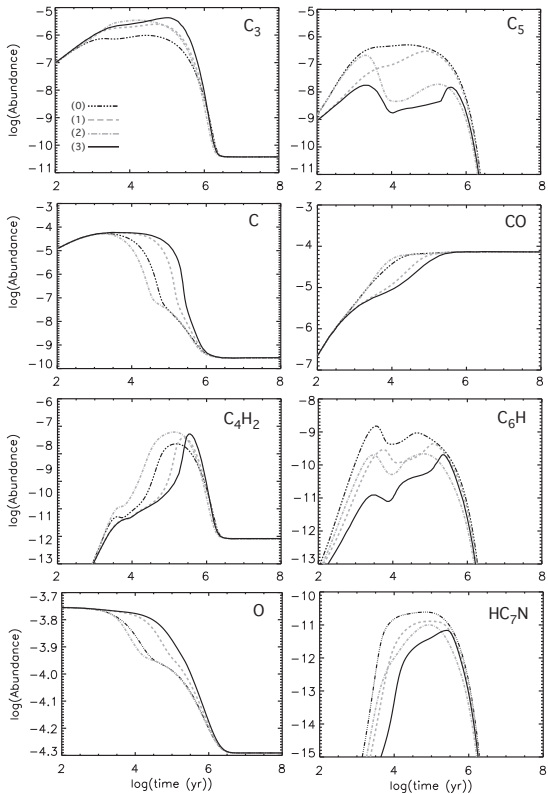}
\caption{Abundances with respect to total hydrogen of several important species. Black  dash-dotted lines (0) are calculated abundances when the standard network osu.2003 is used; grey dashed lines (1) when only C + C$_3$ is modified; grey dash-dotted lines (2) when only C + C$_5$ is modified, and black  solid lines (3) when both C + C$_3$ and C + C$_5$ are modified.  
 \label{SpeAbu}}
\end{figure}

In Fig.~\ref{SpeAbu}, we show how the abundances of selected species are affected by the modifications of the C+C$_3$ and C+C$_5$ reactions at intermediate times; at other times there is little to no effect. For carbon monoxide, the reduction of the C + C$_{3}$ reaction rate coefficient  results in the lowering of its abundance in the period  $10^3$ -- $10^5$~yr by at most one order of magnitude. Changing the products of the C+C$_5$ reaction on the contrary slightly increases the CO abundance.   For neutral atomic carbon, the gradual decrease in abundance with increasing time occurs much later when the rate coefficient of C +C$_{3}$  is reduced; at its maximum, there is a difference in abundance of about 3 orders of magnitude. The change in C + C$_{5}$ speeds up the destruction of C.  In fact, the only species shown in Fig.~\ref{SpeAbu} for which the modification of this reaction has a strong effect are C$_5$ and C$_{6}$H.  

As discussed in Section~\ref{sens}, with the use of the osu.2003 network, these two reactions are both involved in catalytic cycles to convert C and O into CO. By radiative association reactions with atomic carbon, C$_3$ and C$_5$ form C$_4$ and C$_6$, which then react with atomic oxygen to give CO and regenerate C$_3$ and C$_5$.   In this work, we have shown that the C + C$_3$ reaction is less efficient than the value used in osu.2003 by two orders of magnitude at 10 K and the more likely products of the C + C$_5$ reaction are two C$_{3}$ molecules. Even after changing these reactions, this catalytic process  remains the most efficient one to form CO at $10^4$~yr. Changing only the C + C$_5$ reaction products makes this process more efficient by introducing more C$_3$ into the cycle. For this reason, CO is slightly increased whereas C is decreased. Obviously C$_3$ is more abundant because C + C$_3$ consumes less C$_3$ and C + C$_5$ produces it. Contrary to C$_3$, the C$_5$ abundance decreases if  the $\rm C + C_5 \rightarrow C_6$ -- $\rm O + C_6 \rightarrow C_5 + CO$ cycle is broken because C$_5$ is not locked in this loop anymore and is destroyed to give C$_3$. The effect of these changes on C-rich molecules such as the ones shown in Fig.~\ref{SpeAbu} is less evident to explain. From a general point of view,  larger abundances of atomic oxygen and carbon between $10^4$ and $10^5$~yr destroy these species more efficiently than producing them. The radical C$_6$H for instance is destroyed by C to give C$_7$ and C$_7$ in turn reacts with C to produce C$_3$ and C$_5$ (see Table~\ref{CCNrates}). Similarly C$_4$H$_2$ is destroyed by the reaction $\rm C_4H_2 + C \rightarrow C_5H + H$ and C$_5$H is destroyed by O, producing CO and C$_4$H. Large cyanopolyynes, such as HC$_7$N, are reduced in abundance. On the other hand, large O-bearing molecules, such as CH$_3$OH, are increased in abundance  or not changed.

\subsection{Comparison with observations}

\begin{figure}
\includegraphics[width=0.6\linewidth]{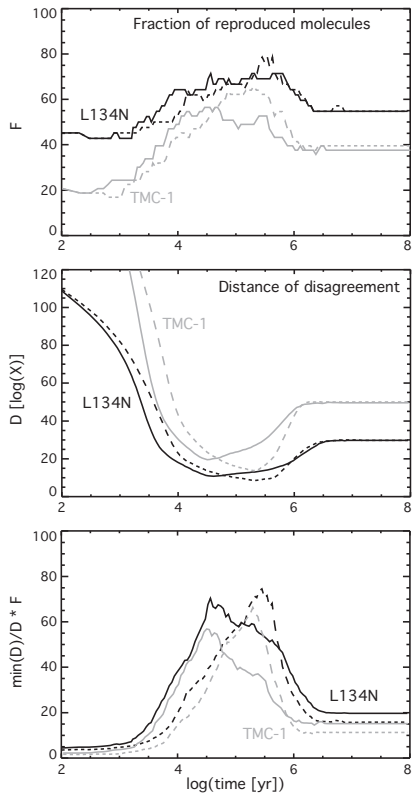}
\caption{Comparison between model and observations in two different clouds (black: L134N and grey: TMC-1). The solid line represents the comparison with the model using the original values of the  C + C$_{\rm n}$ rate coefficients whereas the dashed lines represents the comparisons with the updated model.
 \label{comp_obs}}
\end{figure}

To study the impact of these new rate coefficients on the agreement between models and observations, we used the method described in \citet{2006A&A...451..551W}. We modeled the chemical evolution of two different clouds: L134N and TMC-1 (Cyanopolyyne Peak), using the Nahoon chemical model. The list of reactions is osu.2003 with the original C + C$_{\rm n}$ rate coefficients and osu.2003 with the new values of the C + C$_{n}$ rate coefficients listed in Table~\ref{CCNrates}. For the uncertainties in the rate coefficients of these reactions, we assumed a flat distribution (because the mean values of the rate coefficients are not necessarily the favorite values) with the uncertainty ranges described in Table~\ref{CCNrates}. For the other reactions, we used log-normal distributions \citep[see ][]{2004A&A...422..159W}. The physical conditions are taken to be homogeneous in the observed regions: a gas temperature of 10~K, an H$_2$ density of $10^4$~cm$^{-3}$ and a visual extinction of 10. The elemental abundances are the low-metal elemental abundances \citep[see table 1 in][]{2008ApJ...680..371W}. For TMC-1, we decreased the oxygen abundance to have a C/O elemental ratio of 1.2 in order to better reproduce the observations 
\citep[see ][ and references therein]{2006A&A...451..551W}. The observed abundances are given in Tables 3 and 4 of \citet{2007A&A...467.1103G}. In addition, we included the observational upper limits on the SiO abundances} in these two clouds by \citet{1989ApJ...343..201Z}; the upper limits for this species are $2.6\times 10^{-12}$ and $2.4\times 10^{-12}$ for L134N and TMC-1 respectively.  In total,  we have observed abundances of 42 and 53 species for these two clouds to compare with our calculations.

The results of the comparisons are shown in Fig~\ref{comp_obs} in three panels, for each of which a different criterion is utilized. These criteria are based on the Monte Carlo approach to determination of the uncertainties in the computed abundances based on uncertainties in the rate coefficients, and the overlap of Gaussian distributions representing both the computed and observed abundances \citep{2006A&A...451..551W}.  In the top panel, we use the  fraction of reproduced molecules.  In the middle panel, we use the so-called ``distance of disagreement'' D in which, for species not reproduced by the first criterion, we sum over the remaining species the distances between logarithms of the observed and model fractional abundances.  Finally, in the bottom panel, we use the  minimum distance of disagreement (over all times), divide by the  distance of disagreement, and multiply by the fraction of reproduced molecules. See \citet{2006A&A...451..551W} for details on these parameters.  The time of best agreement is roughly the same for all three parameters, although the sharpest agreement occurs for the bottom panel. It can be seen in all panels that the general agreement with the observations is slightly better using the new values of the  C + C$_{\rm n}$ rate coefficients.  However, the more interesting effect is the shift in time of the best agreement. With the original model, the best agreement was obtained around $3\times 10^4$~yr whereas now we calculate the best agreement to be between $10^5$ and $3\times 10^5$~yr. Note that to compare our results with the results of \citet{2006A&A...451..551W}, we have only updated the C + C$_{\rm n}$ rate coefficients whereas many other reactions have been updated in the latest version of the osu database (osu.3.2008).

\section{Conclusion}
  
This paper presents an approach to improve chemical models, which consists of first identifying important reactions, then obtaining improved rate coefficients for these reactions, and finally 
studying the impact of these new rate coefficients on the predictions of the model.  In this example of the method, we found that under cold dense cloud conditions, the abundances of many species are
sensitive to the rate coefficients of the C + C$_3$ and C + C$_5$ reactions.  Using detailed $ab$ $initio$ and kinetic calculations, we have shown that the rate coefficient of the $\rm C + C_3 \rightarrow C_4$ previously used in the osu.2003 database is incorrect by two to three orders of magnitude at 10 K and has an inverse temperature dependence.  We also found that the rate coefficient for  the  $\rm C + C_5$ reaction was roughly correct in the osu.2003 database, but the products are most likely $\rm C_3 + C_3$ instead of C$_6$, as assumed earlier based on faulty thermodynamic information.  Newly obtained rate coefficients for other reactions in the C + C$_{\rm n}$ series are also reported, although the model shows little sensitivity to these.   With the new rate coefficient for the association reaction between C and C$_{3}$  and the slightly changed rate coefficient and new products for the $\rm C + C_5$ reaction, we found that the abundances of many carbon chain species are strongly modified at the so-called early time of $10^5$~yr. The main reason for this effect is that the $\rm C + C_3 \rightarrow C_4$ / $\rm O + C_4 \rightarrow CO + C_3$ and $\rm C + C_5 \rightarrow C_6$ / $\rm O + C_6 \rightarrow CO + C_5$ reactions comprise a very efficient catalytic cycle to transform C and O into CO.  Slowing down the first cycle and removing the second one, both decreases the abundance of CO and enhances the abundances of atomic C and O at early times.  These changes in turn cause an alteration in the early-time abundances of many other species.  The agreement with observed abundances in the two dense clouds L134N and TMC-1 (CP) is not significantly modified by these new rate coefficients/products but the best agreement is obtained at later times using the new values.  The newly calculated rate coefficients and/or products will be included  in the latest network from the Ohio State group (\verb+http://www.physics.ohio-state.edu/~eric/research.html+). 
  
\begin{acknowledgements}
V. W. acknowledges the french national program PCMI for partial support to this work.
E. H. thanks the National Science Foundation for its support of his research program in astrochemistry. 

\end{acknowledgements}


\bibliographystyle{aa}

\bibliography{aamnem99,biblio}

\begin{thebibliography}{32}
\expandafter\ifx\csname natexlab\endcsname\relax\def\natexlab#1{#1}\fi

\bibitem[{{Bates} \& {Herbst}(1988)}]{1988rcia.conf...17B}
{Bates}, D.~R. \& {Herbst}, E. 1988, in Rate Coefficients in Astrochemistry
  (Dordrecht: Kluwer), ed. T.~J. {Millar} \& D.~A. {Williams}, 41

\bibitem[{{Boy\'e} {et~al.}(2002){Boy\'e}, {Campos}, {Douin}, {Follows},
  {Gauyacq}, {Shafizadeh}, {Halvick}, \& {Bowers}}]{2002Boye}
{Boy\'e}, S., {Campos}, A., {Douin}, S., {et~al.} 2002, J. Chem. Phys., 116,
  8843

\bibitem[{{Chastaing} {et~al.}(2001){Chastaing}, {Le Picard}, {Sims}, \&
  {Smith}}]{2001A&A...365..241C}
{Chastaing}, D., {Le Picard}, S.~D., {Sims}, I.~R., \& {Smith}, I.~W.~M. 2001,
  A\&A, 365, 241

\bibitem[{{Choi} {et~al.}(2000){Choi}, {Bise}, {Hoops}, {Mordaunt}, \&
  {Neumark}}]{Choi2000}
{Choi}, H., {Bise}, R.~T., {Hoops}, A.~A., {Mordaunt}, D.~H., \& {Neumark},
  D.~M. 2000, J. Phys. Chem. A, 104, 2025

\bibitem[{{Clary} {et~al.}(1994){Clary}, {Haider}, {Husain}, \&
  {Kabir}}]{1994ApJ...422..416C}
{Clary}, D.~C., {Haider}, N., {Husain}, D., \& {Kabir}, M. 1994, ApJ, 422, 416

\bibitem[{{Dunning}(1989)}]{1989Dunning}
{Dunning}, T.~H. 1989, J. Chem. Phys., 90, 1007

\bibitem[{{Forst}(1993)}]{1993Forst}
{Forst}, W. 1993, Quantum Chem. Program. Exch., 13, 21

\bibitem[{{Garrod} {et~al.}(2007){Garrod}, {Wakelam}, \&
  {Herbst}}]{2007A&A...467.1103G}
{Garrod}, R.~T., {Wakelam}, V., \& {Herbst}, E. 2007, \aap, 467, 1103

\bibitem[{{Gingerich} {et~al.}(1994){Gingerich}, {Finkbeiner}, \&
  {Schmude}}]{Gingerich1994}
{Gingerich}, K.~A., {Finkbeiner}, H.~C., \& {Schmude}, W.~J. 1994, J. Am. Chem.
  Soc., 116, 3884

\bibitem[{{Herbst}(1982)}]{1982Herbst}
{Herbst}, E. 1982, Chem. Phys., 65, 185

\bibitem[{{Herbst} \& {Bates}(1988)}]{1988ApJ...329..410H}
{Herbst}, E. \& {Bates}, D.~R. 1988, ApJ, 329, 410

\bibitem[{{Holbrook} {et~al.}(1996){Holbrook}, {Pilling}, \&
  {Roberston}}]{1996Holbrook}
{Holbrook}, K.~A., {Pilling}, M.~J., \& {Roberston}, S.~H. 1996, {Unimolecular
  Reactions. 2nd Ed.} (John Wiley \& Sons Ltd.: Chichester)

\bibitem[{{Larsson}(1983)}]{1983Larson}
{Larsson}, M. 1983, ApJ, 128, 291

\bibitem[{{Linnart} {et~al.}(2000){Linnart}, {Vaizert}, {Motylewsky}, \&
  {Maier}}]{2000Linnartz}
{Linnart}, H., {Vaizert}, O., {Motylewsky}, T., \& {Maier}, J.~P. 2000, J.
  Chem. Phys., 112, 9777

\bibitem[{{Martin} \& {Taylor}(1995)}]{Martin1995}
{Martin}, J.~M.~L. \& {Taylor}, P.~R.~J. 1995, J. Chem. Phys., 102, 8270

\bibitem[{{Masso} {et~al.}(2006){Masso}, {Senent}, {Rosmus}, \&
  {Hochlaf}}]{2006Masso}
{Masso}, H., {Senent}, M.~L., {Rosmus}, P., \& {Hochlaf}, M. 2006, J. Chem.
  Phys., 124, 234304

\bibitem[{{Mebel} {et~al.}(2007){Mebel}, {Kislov}, \& {Hayashi}}]{2007Mebel}
{Mebel}, A.~N., {Kislov}, V.~V., \& {Hayashi}, M. 2007, J. Chem. Phys., 126,
  204310

\bibitem[{{Nelson} {et~al.}(1982){Nelson}, {Helvajian}, {Pasternack}, \&
  {McDonald}}]{1982Nelson}
{Nelson}, H.~H., {Helvajian}, H., {Pasternack}, L., \& {McDonald}, J.~R. 1982,
  Chem. Phys., 73, 431

\bibitem[{{Press} \& {Flannery}(1992)}]{NumericalRecipes}
{Press}, W.~H. \& {Flannery}, B.~P. 1992, {Numerical Recipes} (Cambridge
  University Press, Cambridge)

\bibitem[{{Quan} {et~al.}(2008){Quan}, {Herbst}, {Millar}, E., {Lin}, {Guo},
  {Honvault}, \& {Xie}}]{quan2008}
{Quan}, D., {Herbst}, E., {Millar}, T.~J., {et~al.} 2008, ApJ, 681, 1318

\bibitem[{{Smith}(1989)}]{1989Smith}
{Smith}, I.~W.~M. 1989, Chem. Phys., 131, 391

\bibitem[{{Smith} {et~al.}(2004){Smith}, {Herbst}, \&
  {Chang}}]{2004MNRAS.350..323S}
{Smith}, I.~W.~M., {Herbst}, E., \& {Chang}, Q. 2004, MNRAS, 350, 323

\bibitem[{{Smith} {et~al.}(2006){Smith}, {Sage}, {Donahue}, {Herbst}, \&
  {Park}}]{2006smith}
{Smith}, I.~W.~M., {Sage}, A.~M., {Donahue}, N.~M., {Herbst}, E., \& {Park},
  I.~H. 2006, in { Chemical Evolution of the Universe. Faraday Discussions No.
  133}, 137--156

\bibitem[{{Vasyunin} {et~al.}(2008){Vasyunin}, {Semenov}, {Henning}, {Wakelam},
  {Herbst}, \& {Sobolev}}]{2008ApJ...672..629V}
{Vasyunin}, A.~I., {Semenov}, D., {Henning}, T., {et~al.} 2008, ApJ, 672, 629

\bibitem[{{Vasyunin} {et~al.}(2004){Vasyunin}, {Sobolev}, {Wiebe}, \&
  {Semenov}}]{2004AstL...30..566V}
{Vasyunin}, A.~I., {Sobolev}, A.~M., {Wiebe}, D.~S., \& {Semenov}, D.~A. 2004,
  Astronomy Letters, 30, 566

\bibitem[{{Wakelam} {et~al.}(2004){Wakelam}, {Caselli}, {Ceccarelli}, {Herbst},
  \& {Castets}}]{2004A&A...422..159W}
{Wakelam}, V., {Caselli}, P., {Ceccarelli}, C., {Herbst}, E., \& {Castets}, A.
  2004, A\&A, 422, 159

\bibitem[{{Wakelam} \& {Herbst}(2008)}]{2008ApJ...680..371W}
{Wakelam}, V. \& {Herbst}, E. 2008, ApJ, 680, 371

\bibitem[{{Wakelam} {et~al.}(2006{\natexlab{a}}){Wakelam}, {Herbst}, \&
  {Selsis}}]{2006A&A...451..551W}
{Wakelam}, V., {Herbst}, E., \& {Selsis}, F. 2006{\natexlab{a}}, A\&A, 451, 551

\bibitem[{{Wakelam} {et~al.}(2006{\natexlab{b}}){Wakelam}, {Herbst}, {Selsis},
  \& {Massacrier}}]{2006A&A...459..813W}
{Wakelam}, V., {Herbst}, E., {Selsis}, F., \& {Massacrier}, G.
  2006{\natexlab{b}}, A\&A, 459, 813

\bibitem[{{Wakelam} {et~al.}(2005){Wakelam}, {Selsis}, {Herbst}, \&
  {Caselli}}]{2005A&A...444..883W}
{Wakelam}, V., {Selsis}, F., {Herbst}, E., \& {Caselli}, P. 2005, A\&A, 444,
  883

\bibitem[{{Xu} {et~al.}(1997){Xu}, {Burton}, {Taylor}, \& {Neumark}}]{1997Xu}
{Xu}, C., {Burton}, G.~R., {Taylor}, T.~R., \& {Neumark}, D.~M. 1997, J. Chem.
  Phys., 107, 3428

\bibitem[{{Ziurys} {et~al.}(1989){Ziurys}, {Friberg}, \&
  {Irvine}}]{1989ApJ...343..201Z}
{Ziurys}, L.~M., {Friberg}, P., \& {Irvine}, W.~M. 1989, ApJ, 343, 201

\end{thebibliography}



\appendix
\section{Other $\rm C + C_n$ reactions}

The calculated rate coefficients and products for C + C$_{\rm n}$ are listed in Table~\ref{CCNrates} for $\rm n = 2-8$ along with estimated uncertainties and the older values used in the osu.2003 network.  The applicable temperature range is generally 10-300 K, although for radiative association reactions, the dependence can extend to higher temperatures.   For the C($\rm ^3P$) + C$_2$($\rm ^1\Sigma ^+_g$) reaction, the spin-allowed pathway leads to a C$_3$($\rm a~^3\Pi _u$) adduct,  which is not the ground electronic state of C$_3$ but is highly thermodynamically accessible, lying  5.16 eV below the reactants.  Preliminary $ab$ $intio$ calculations (CASCCF, vdz basis, full valence active space) give no barrier for the production of the $\rm a^3\Pi _u$ state. To determine the rate coefficient for radiative association, we considered only vibrational relaxation,  Computing vibrational frequencies for C$_{3}$ at the B3LYP/vtz level, we obtained a relaxation rate of 
$k_{\rm rad,\nu}$  = 220 s$^{-1}$.  We then estimated the association (capture) rate coefficient to be  $2 \times 10^{-10}$ cm$^3$ s$^{-1}$ based on experiments of analogous systems (e.g. C + C$_{2}$H$_{2}$)  and calculated the dissociation rate coefficient by the RRKM method.   We obtained the radiative association rate coefficient, shown in Table~\ref{CCNrates}, to be  $(3 \pm 2) \times 10^{-16} \times 
(T/300)^{-1.0}$ cm$^3$ s$^{-1}$, a rather low value due to the rapid dissociation of the small C$_{3}$ adduct. 

For the C($\rm ^3P$) + C$_4$($\rm ^3\Sigma^-_g$), C$_6$($\rm ^3\Sigma^-_g$), C$_8$($\rm ^3\Sigma^-_g$) reactions, the formation of the C$_{\rm n+1}$ adducts is strongly exothermic and spin-allowed.  The possible potential energy surfaces in the entrance channels for all of these reactions are $^{1,3,5} \Sigma^+$ and $^{1,3,5} \Pi$.  Preliminary $ab$ $intio$ calculations (CASCCF, vdz basis) show that the three $^{1,3,5} \Sigma ^+$ states have no barrier in the entrance channel, but that the three $^{1,3,5} \Pi$ states have high barriers along the reaction pathway.  Exothermic fragment channels exist for all of these adducts.  The rate coefficients for these reactions are once again estimated from experimental values of analogous systems, which in general are smaller than the capture values, derived from the formalism of \citet{1994ApJ...422..416C}, and further reduced by a factor that allows for the multiplicity of the potential energy surfaces that correlate with the reagents and also by a factor for the multiplicity of the exit channels if relevant.   Correlation involving the fine-structure states of atomic C is ignored. The assigned  rate coefficients and estimated uncertainties are shown in Table~\ref{CCNrates}.

The C($\rm ^3P$) + C$_7$($\rm ^1\Sigma^+_g$) reaction leading to the formation of the C$_8$ adduct in the $\rm ^3\Sigma^-_g$ ground electronic state is strongly exothermic and spin-allowed. By comparison with C + C$_5$ we assume that there is no barrier in the entrance channel for the formation of the ground C$_8$ electronic state and, although this reaction is formally spin-forbidden to the products in their singlet ground states, the adduct is once again likely to last long enough to allow intersystem crossing.  Once again, we use analogous experimental results rather than the  formalism of \citet{1994ApJ...422..416C} to obtain the capture rate, and reduce this via eq.~(\ref{finestruc}).  

\begin{table}
\caption{Rate coefficients for the $\rm C + C_n$ reactions. }
\begin{tabular}{|l|c|c|c|c|}
\hline
\hline
Reaction & \multicolumn{2}{|c|}{osu.2003 rate coefficient} & \multicolumn{2}{|c|}{New rate coefficient} \\
 & $\alpha^{1}$ &  $\beta^{1}$ & $\alpha$ &  $\beta$ \\
\hline
$\rm C + C_2 \longrightarrow C_3 + h\nu$ & -- & -- & $(3 \pm 2) \times 10^{-16}$ & -1 \\
$\rm C + C_3 \longrightarrow C_4 + h\nu$ & $1.0\times 10^{-10}$ & 0 & $(4\pm 3)\times 10^{-14}$ & -1 \\
$\rm C + C_4 \longrightarrow C_2 + C_3$ & $1.0\times 10^{-13}$ & 0 & $(0.9 \pm 0.6)\times 10^{-10}$ & 0 \\
$\rm C + C_5 \longrightarrow C_6 + h\nu$ & $1.0 \times 10^{-10}$ & 0 & -- & -- \\
$\rm C + C_5 \longrightarrow C_3 + C_3$ & -- & -- & $(1.2\pm 0.8)\times 10^{-10}$ & 0 \\
$\rm C + C_6 \longrightarrow C_2 + C_5$ & $1.0\times 10^{-13}$ & 0 & $(0.5\pm 0.3) \times 10^{-10}$ & 0 \\
$\rm C + C_6 \longrightarrow C_3 + C_4$ & -- & -- & $(0.5\pm 0.3) \times 10^{-10}$ & 0 \\
$\rm C + C_7 \longrightarrow C_8 + h\nu$ & $1.0 \times 10^{-10}$ & 0 & -- & -- \\
$\rm C + C_7 \longrightarrow C_3 + C_5$ & -- & -- & $(1.4 \pm 0.8) \times 10^{-10}$ & 0 \\
$\rm C + C_8 \longrightarrow C_4 + C_5$ & -- & -- & $(0.4 \pm 0.2)\times 10^{-10}$ & 0 \\
$\rm C + C_8 \longrightarrow C_3 + C_6$ & -- & -- & $(0.4 \pm 0.2)\times 10^{-10}$ & 0 \\
$\rm C + C_8 \longrightarrow C_2 + C_7$ &  $1.0 \times 10^{-13}$ & -- & $(0.4 \pm 0.2)\times 10^{-10}$ & 0 \\
\hline
\end{tabular}
\label{CCNrates}
\\
$^1$ Rate coefficients in units of cm$^{3}$ s$^{-1}$ given by $k = \alpha \times (T/300)^{\beta}$ \\
\end{table}%

\end{document}